\documentclass[reprint,
 amsmath,
 amssymb,
 floatfix]{revtex4-1}

\usepackage{amsmath,amssymb,amsfonts}
\usepackage[normalem]{ulem}
\usepackage[usenames, dvipsnames]{color}
\usepackage{graphicx}
\usepackage{dcolumn}
\usepackage[caption=false]{subfig}
\usepackage{hyperref} 
\hypersetup{colorlinks=true} 

\pdfoutput=1

\begin{document}

\title{Transient Synchronisation and Quantum Coherence in a Bio-Inspired Vibronic Dimer}

\author{Stefan Siwiak-Jaszek}
 \email{stefan.siwiak-jaszek.11@ucl.ac.uk}

\author{Alexandra Olaya-Castro}%
 \email{a.olaya@ucl.ac.uk}
\affiliation{Department of Physics and Astronomy, University College London, London WC1E 6BT, United Kingdom}%

\date{13 January 2019}

\begin{abstract}
Synchronisation is a collective phenomenon widely investigated in classical oscillators and, more recently, in quantum systems. However, it remains unclear what features distinguish synchronous behaviour in these two scenarios. Recent works have shown that investigating the dynamics of synchronisation in open quantum systems can give insight into this issue. Here we study transient synchronisation in a bio-inspired vibronic dimer, where the dynamics of electronic excitation is mediated by coherent interactions with intramolecular vibrational modes. We show that the synchronisation dynamics of the displacement of these local modes exhibit a rich behaviour which arises directly from the distinct time-evolutions of different vibronic quantum coherences. Furthermore, our study shows that coherent energy transport in this bio-inspired system is concomitant with the emergence of positive synchronisation between mode displacements. Our work provides further understanding of the relations between quantum coherence and synchronisation in open quantum systems and suggests an interesting role for coherence in biomolecules, that is promoting the synchronisation of vibrational motions driven out of thermal equilibrium.
\end{abstract}

\maketitle

\section{\label{sec:intro}Introduction}
Synchronisation can be broadly defined as the adjustment of rhythms of oscillating objects due to their weak interaction \cite{Pikovsky2001}. This description corresponds to numerous processes throughout the natural world that occur on a wide range of length and time scales. This is especially true in living systems, where synchronisation is common and is often closely related to biological function \cite{Buck1966, Nitsan2016}. On the metre scale, the synchronous flashing of male fireflies aids each individual in reproduction \cite{Buck1966} and on the micrometre scale, cardiac cells synchronise contractions with their neighbours \cite{Nitsan2016}. The question of whether this phenomenon persists on even smaller length scales such as that of individual biomolecules (nanometre) and time scales such as the relaxation times of intramolecular motions (picoseconds) has not been investigated. On these length and time scales quantum phenomena cannot be neglected and synchronisation, if it occurs, could exhibit features that have no equivalence in the classical regime \cite{Lorch2017,Giorgi2016,Qiu2015,Hush2015,Walter2014a,Lee2013}.

An interesting biophysical scenario in which one can explore synchronisation in the quantum regime, and at the same time investigate its possible relations to coherence and biological function, is during electronic excitation transport in photosynthetic complexes \cite{Fassioli2014,Scholes2017}, where coherent dynamics has been observed lasting several hundred femtoseconds \cite{Engel2007,Collini2010,Harel2011,Panitchayangkoon2011,Romero2014a, Fuller2014}. The leading hypothesis for the mechanism underlying long-lived coherent dynamics in these biophysical systems is the quantum mechanical exchange of energy between excitonic and vibrational degrees of freedom \cite{Kolli2012, Richards2012, Chin2013, Huelga2013, OReilly2014, Romero2014a, Fuller2014, Novelli2015,Chen2016,Dean2016}. Despite some controversy surrounding the observations \cite{Duan2017}, there remains a widespread interest in understanding the intertwined dynamics of electronic and vibrational motions during energy transfer. In this context, it is timely and interesting to investigate whether molecular vibrations may be synchronised during ultra-fast energy transfer processes in photosynthetic complexes and, if such synchronisation happens, what its relations to coherence and function could be.

The form of synchronisation that we study here should be described as \textit{transient}, emphasising that it occurs before relaxation to the ground state, and \textit{spontaneous}, meaning it arises due to the interactions within the quantum system considered and not due to the influence of an external fixed-frequency driving force \cite{Giorgi2012}. We note that this differs from other investigations of synchronisation in self-sustained quantum oscillators \cite{Walter2014a,Lorch2017}. 

Transient spontaneous quantum synchronisation (referred to from here on as synchronisation) has been recently investigated in a range of open quantum systems \cite{Giorgi2012,Giorgi2013,Benedetti2016,Militello2017,Manzano2013}. In the simplest case of coupled two-level systems (TLSs), it has been shown that synchronisation cannot occur in the presence of dephasing channels alone and dissipation appears to be essential \cite{Giorgi2013}. In quantum harmonic oscillator (QHO) networks, adjusting the coupling between oscillators can create asymptotically synchronised states that avoid dissipation altogether \cite{Manzano2013}; and in hybrid TLS-QHO systems the form of coupling of QHOs to TLSs can induce and control synchronisation between QHOs with a well-defined phase difference \cite{Militello2017}. From these works it can be concluded that the specific forms of the decoherence channels and of the interactions among quantum subsystems play a pivotal role in reaching synchronisation. However, exactly how the interplay between coherent dynamics and decoherence enables synchronisation in a particular time scale, or how coherences may relate to synchronisation is not fully understood. Here, we show that the investigation of synchronisation in a bio-inspired vibronic dimer provides interesting insights on these open questions.

In particular, we investigate the synchronisation of molecular motions during electronic energy transport in a prototype photosynthetic vibronic dimer where two local excited electronic states interact with each other and with local quasi-coherent intramolecular modes. The modes are subject to local dissipation into thermal baths while the electronic subsystem undergoes pure dephasing. We provide an intuitive explanation of the mechanism behind synchronisation by analysing the eigenstate coherences that influence the oscillatory patterns of the local positions being synchronised. Our results indicate that interferences between vibronic coherences can result in negative and positive synchronisation. The periods of negative and positive synchronisation reflect a rich and distinctive interplay between coherence and decoherence mechanisms at different timescales. Furthermore, we show that a faster onset of synchronisation is correlated with a larger degree of coherent excitation transport. From a biological standpoint, our study suggests local mode synchronisation may be present during energy transfer in some photosynthetic pigment-protein complexes at physiological temperatures.

\section{\label{sec:methods}Modelling synchronisation in exciton-vibration dimers}
In the following sections we describe the model and methods used. In Section \ref{sec:exciton-vibration dimer} we introduce the Hamiltonian for the exciton-vibration dimer model that can represent dimers in a variety of photosynthetic pigment-protein complexes. In Section \ref{sec:Open Systems Model} we describe the Markovian master equation that phenomenologically describes the open quantum system dynamics of the exciton-vibration dimer and briefly discuss our numerical methods used to solve it. In Section \ref{sec:synch measure} we describe the measure used to quantify synchronisation and discuss its limitations.

\subsection{The Exciton-Vibration Dimer} \label{sec:exciton-vibration dimer}
\begin{figure}[ht]
\centering
\includegraphics[width=\textwidth,height=5cm,keepaspectratio]{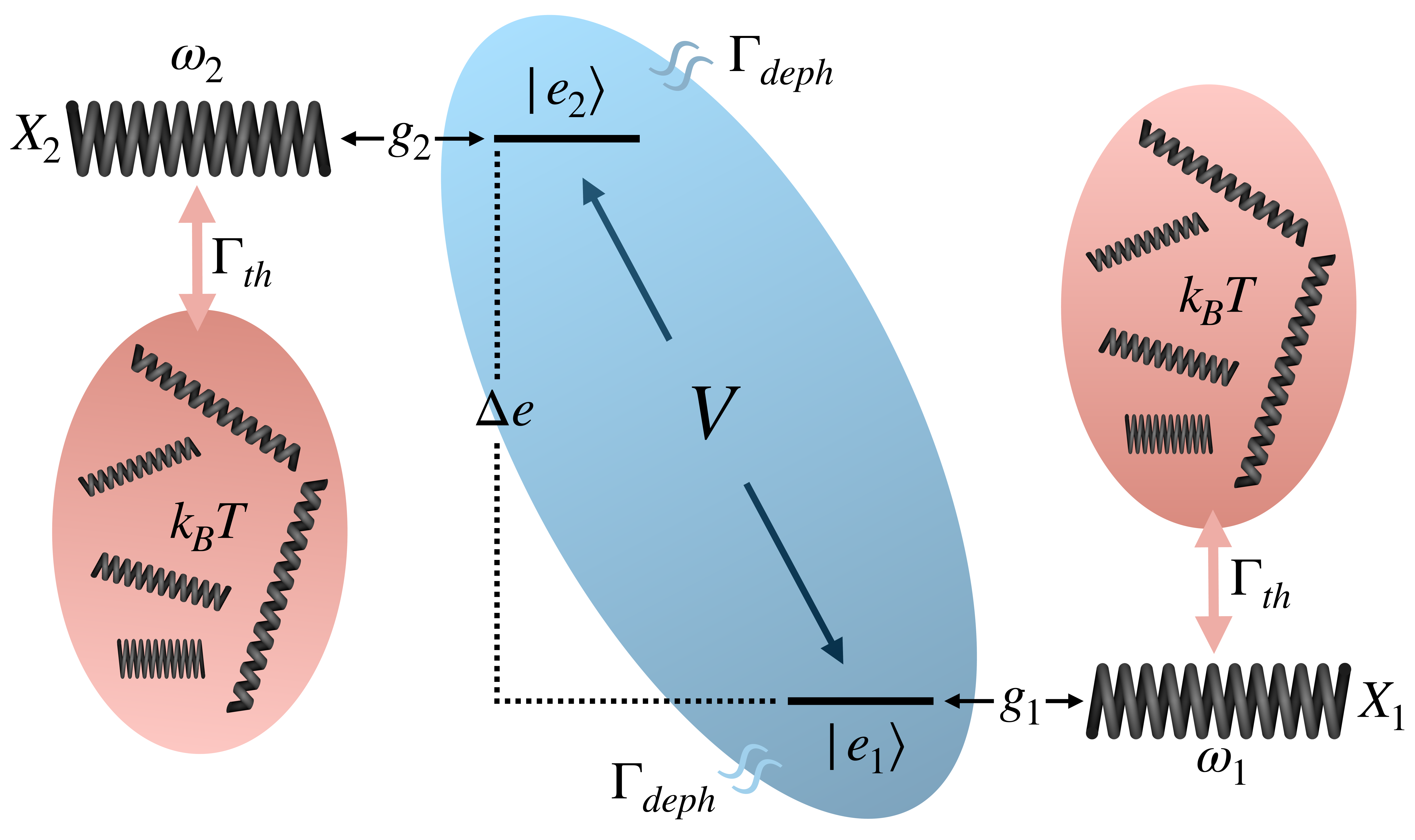}
\caption{Schematic diagram of the exciton-vibration dimer. Two chromophores (subscripts 1 and 2) with single excited states $|e_i\rangle$ interacting via dipole-dipole coupling of strength $V$. Each electronic state is coupled linearly with strength $g$ to a harmonic mode of energy $\omega$. Electronic sub-system (blue oval) experiences pure dephasing of rate $\Gamma_{deph}$, each mode dissipates into separate thermal baths (red ovals) of temperature $k_BT$ at rate $\Gamma_{th}$.}
\label{Model}
\end{figure}

Photosynthetic light-harvesting proteins exhibit complex excitation transfer (ET) dynamics due to an overlap of energy scales in electronic and vibrational degrees of freedom and the timescales of their associated coherent and incoherent processes \cite{Fassioli2014}. Experiments and theory suggest that these complexes are capable of sustaining quantum coherence at room temperature that lasts several hundred femtoseconds, thereby suggesting a functional role for coherent dynamics \cite{Fassioli2014, Scholes2017}. Here we consider a prototype light-harvesting unit, which we call an exciton-vibration dimer model, formed by a pair of chromophores whose local electronic excitations interact with quasi-coherent vibrational modes \cite{Kolli2012,OReilly2014}.

Let the chromophores have single excited states $|e_{i=1,2}\rangle$ of energy $e_{i=1,2}$ which interact via dipole-dipole coupling of strength $V$ and are both locally coupled to a quantised intramolecular mode of energy $\omega_{i=1,2}$ with strength $g_{i=1,2}$. The system has a total Hamiltonian of the form: $H=H_{el}+H_{vib}+H_{el-vib}$. The electronic Hamiltonian reads:
\begin{equation}
    H_{el} = e_1 |e_1\rangle \langle e_1| + e_2 |e_2\rangle \langle e_2| + V|e_2\rangle \langle e_1|+V^*|e_1\rangle \langle e_2| 
\end{equation}
and its eigenstates are delocalised electronic states known as excitons $|E_{d=1,2}\rangle$ with energies:
\begin{equation}
    \begin{split}
    E_{1} = & \frac{1}{2}\left(e_1+e_2 - \sqrt[]{\Delta e^2 + 4|V|^2}\right) \\
    E_{2} = & \frac{1}{2}\left(e_1+e_2 + \sqrt[]{\Delta e^2 + 4|V|^2}\right),
    \end{split}
\end{equation}
where $\Delta e = e_2-e_1$ and $|V|= \sqrt{VV^*}$.

The unitary rotation $U(\theta)$ which diagonalises $H_{el}$ to create an excitonic Hamiltonian $H_{exc}=U(\theta)H_{el}U^\dag(\theta)=E_1|E_1\rangle\langle E_1| + E_2|E_2\rangle\langle E_2|$ has the form \cite{MayKuhnBook2011}:
\begin{equation}
    U = \begin{pmatrix}\cos\theta & \sin\theta \\ -\sin\theta & \cos\theta\end{pmatrix}
\end{equation}
where $\theta = \frac{1}{2}\arctan(2|V|/\Delta e)$ is referred to as the mixing angle and is an effective measure of delocalisation of the electronic subsystem or the exciton size. 

The bare vibrational Hamiltonian reads:
\begin{equation}
    H_{vib} = \omega_1 b^\dag_1 b_1 + \omega_2 b^\dag_2 b_2 
\end{equation}
where $b_{i=1,2}^\dag(b_{i=1,2})$ are the creation (annihilation) operators for the modes. The eigenstates of $H_{vib}$ are fock states which we write as: $|n_1\rangle \otimes |n_2\rangle$ where $n$ are the fock state numbers and subscripts indicate the mode subspace.

The interaction Hamiltonian reads $H_{el-vib} = g_1 |e_1\rangle \langle e_1|(b_1+b^\dag_1) + g_2 |e_2\rangle \langle e_2|(b_2+b^\dag_2)$ where we have assumed a linear coupling between mode position and electronic states \cite{Renger2004,Schatz2002}. We define the electronic operators in the exciton basis as $\Theta_i = U(\theta)|e_i\rangle \langle e_i|U^\dag(\theta)$ and insert into $H_{el-vib}$ to give:
\begin{equation}
    H_{exc-vib} = g_1 \Theta_1(b_1+b^\dag_1) + g_2 \Theta_2(b_2+b^\dag_2)
\end{equation}

The final exciton-vibration Hamiltonian is then:
\begin{equation}
    \begin{split}
    H = & E_1|E_1\rangle\langle E_1| + E_2|E_2\rangle\langle E_2| \\ & + \omega_1 b^\dag_1b_1 + \omega_2 b^\dag_2b_2 \\ & + g_1 \Theta_1 X_1 + g_2 \Theta_2X_2
    \end{split}
    \label{System Hamiltonian}
\end{equation}
where we have introduced the position operator for each mode $X_{i=1,2} = b_i+b_i^\dag$. The eigenstates of H are exciton-vibrational which we can represent in the local basis as:
\begin{equation}
    \begin{split}
    |\psi_j\rangle = & \sum_{d=1,2}\alpha_d|E_d\rangle \otimes \sum^{M}_{n_1=1}\beta_{n_1} |n_1\rangle\otimes  \sum^{M}_{n_2=1}\gamma_{n_2} |n_2\rangle \\ = & \sum_{d,n_1,n_2}c(d,n_1,n_2)|E_d,n_1,n_2\rangle
    \end{split}
\end{equation}
where eigenstates $|\psi_j\rangle$ are labelled in ascending energy. To obtain convergent dynamics we account for a maximum occupation $M=$ 8 in each mode. 

In this paper we investigate the synchronisation of oscillations in the expectation value of the position operator $\langle X_i \rangle$ for each mode, which we refer to as local mode displacements. We assume equal frequencies $\omega_1 = \omega_2 = \omega$ and identical coupling strengths $g_1=g_2 =g$. We also consider the regime of weak electronic coupling where $\Delta E\approx\omega>g>V$ which is characteristic of chromophore pairs present in a variety of light-harvesting proteins \cite{Womick2011, Richards2012, Kolli2012,Viani2014,Doust2004,Novoderezhkin2010,Novelli2015,Collini2010}. In this regime excitons are not fully delocalised and excitonic energies are close to the local energies. The resultant quasi-localised nature of eigenstates gives validity to our analysis of synchronisation of local mode displacements.

\subsection{Open Quantum System Model}\label{sec:Open Systems Model}
For simplicity, we assume Markovian relaxation processes described in the Lindblad form:
\begin{equation} \label{eq: master}
    \dot{\rho}(t) = -i[H,\rho(t)] + D_{deph}[\rho(t)] + D_{th}[\rho(t)],
\end{equation}
where $\rho(t)$ is the exciton-vibration density matrix and the Lindblad-form superoperators $D_{\nu}[\rho] = \Gamma_{\nu}\left(O_{\nu} \rho O^\dag_{\nu} - \frac{1}{2}\rho O_{\nu}^\dag O_{\nu} - \frac{1}{2}O^\dag_{\nu} O_{\nu}\rho\right)$ for the incoherent operator  $O_\nu$  at rate $\Gamma_{\nu}$. We assume local pure dephasing processes on the electronic sub-system \cite{Haken1973,Breuer2002} with operators $|e_1\rangle\langle e_1|$ and $|e_2\rangle\langle e_2|$ at equal rates of $\Gamma_{deph}=$[0.1 ps]$^{-1}$ such that exciton coherence decays in approximately 0.5 ps as inferred from experimental evidence of algal photosynthetic protein PC645 \cite{Richards2012}. Each mode is assumed to undergo relaxation \cite{Breuer2002} due to thermal reservoirs at temperature 298K (207.1 cm$^{-1}$) which is represented by transition operators $b_1$ and $b_2$ at rate $\Gamma_{th}(1+B)$ and $b^\dag_1$ and $b^\dag_2$ at rate $\Gamma_{th}B$. Here $B = (e^{\frac{\omega}{k_BT}}-1)^{-1}$ is the mean number of quanta in a thermally occupied mode of frequency $\omega$ and $\Gamma_{th}=$[1 ps]$^{-1}$ is the rate at which modes equilibrate. Table \ref{numerical parameters} summarises the parameters considered.

To proceed numerically we linearise the master equation:
\begin{equation}
    |\dot{\rho}(t)\rangle \mkern-3mu \rangle = \mathcal{L}|\rho(t) \rangle \mkern-3mu \rangle
\end{equation}
where $\mathcal{L}$ is the Liouvillian superoperator and $|\rho(t) \rangle \mkern-3mu \rangle$ are flattened density matrices. We solve this ordinary differential equation in Python 3 with packages NumPy and SciPy and author generated scripts.

\begin{table}
\begin{center}
\begin{ruledtabular}
\begin{tabular}{c|c|c|c|c|c|c}
$\Delta e$&$V$&$\omega$&$g$&$k_BT$&$\Gamma_{th}$&$\Gamma_{deph}$\\
\hline
1042&92&1111&267.1&207.1&[1ps]$^{-1}$&[0.1ps]$^{-1}$
\end{tabular}
\end{ruledtabular}
\end{center}
\caption{Parameters used for numerical calculations representing the central dimer in the cryptophyte antennae PE545 ($PEB_{50/61}$) \cite{Novoderezhkin2010,Kolli2012}. All units in spectroscopic wavenumbers cm$^{-1}$ except for the final two columns which are specified in table.}
\label{numerical parameters}
\end{table}

Since we are interested in the process of synchronisation during energy transfer, we fix the initial electronic state to be the higher energy excitonic state $|E_2\rangle$ and assume that both intramolecular modes are initially in thermal equilibrium with their respective baths. This results in the initial state:
\begin{equation}
    \rho(0) = |E_2\rangle\langle E_2|\otimes \rho^{th}_1\otimes\rho^{th}_2
    \label{eqn: initial state}
\end{equation}
where $\rho_{i}^{th}=\sum_{n_i}P_{n_i}|n_i\rangle\langle n_i|$ and $P_{n_i} = \left(1-e^{\frac{-\omega}{k_BT}}\right) e^{\frac{-n_i \omega}{k_BT}}$.

\subsection{Measuring Synchronisation}\label{sec:synch measure}

\begin{figure}[ht]
\centering
\includegraphics[width=0.9\columnwidth]{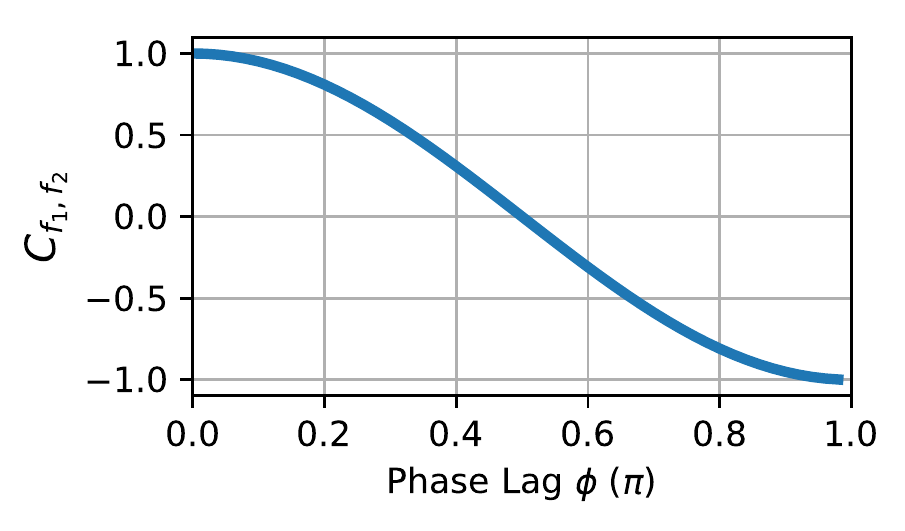}
\caption{Value of synchronisation measure $C_{f_1,f_2}$ for two identical sinusoids as a function of their phase shift $\phi$. $f_1 = \sin (at)$, $f_2=\sin(at+\phi)$, $\Delta t = 1/a$.}
\label{synch function characterisation}
\end{figure}

To quantify synchronisation between signals we adopt a commonly used measure based on the Pearson correlation factor \cite{Galve2016}. Defined generally for any two time dependent functions $f_1(t)$ and $f_2(t)$:
\begin{equation}
    C_{f_1,f_2}(t|\Delta t) = \frac{\int_{t}^{t+\Delta t} \delta f_1 \delta f_2 dt}{\big(\int_{t}^{t+\Delta t}\delta f_1^2 dt \int_{t}^{t+\Delta t}\delta f_2^2 dt\big)^{1/2}}
    \label{correlationcoeff}
\end{equation}
where $\delta f = f - \bar{f}$, $\bar{f} = \frac{1}{\Delta t} \int_{t}^{t+\Delta t} f(t') dt'$ is a time average and $\Delta t$ is the averaging window. 

This function returns a continuous value in the range of -1 to 1 corresponding to a phase shift between $f_1$ and $f_2$ of $\pi$ to $0$. We choose this function for its time dependent quantitative measure of synchronisation and for its wide use in the quantum synchronisation literature \cite{Giorgi2013,Manzano2013,Benedetti2016,Bellomo2017}. An analysis of different synchronisation measures \cite{Galve2016} states that the Pearson correlation factor returns a value of 1 for positive synchronisation (in-phase), -1 for negative synchronisation ($\pi$ out of phase) and 0 for asynchrony. Here however we note that, by an appropriate choice of $\Delta t$ to be as close as possible to the time period of the dominant frequency in $f_{1,2}(t)$, the synchronisation measure in fact indicates the phase difference between two oscillating signals. Therefore any constant value of $C_{f_1,f_2}(t)$ corresponds to a constant phase between the signals e.g. $C_{f_1,f_2}(t)=0$ corresponds to a phase difference of $\pi/2$ at time $t$. A characterisation of this relationship is presented in Figure \ref{synch function characterisation} where the value of the synchronisation function is plotted as a function of constant phase difference, $\phi$, between two perfect sinusoids. Based on this, we define our condition for a synchronised state as a constant value in $C_{f_1,f_2}(t)$ over time.

\section{Results}\label{sec:Results}
In the following sections we present and explain the main findings of this paper. In Section \ref{sec: closed dynamics} we introduce our exciton-vibration coherences description of synchronisation by considering the purely coherent dynamics of the dimer. In Section \ref{Sec:Spontaneous Synch} we show that when dissipative processes are included, spontaneous synchronisation of the displacements of intramolecular modes in exciton-vibration dimers emerges and it is accompanied by a negatively synchronised transient. We illustrate how this synchronisation can be understood as the dominance of a specific exciton-vibration coherence over a set of competing coherences that contribute to oscillatory dynamics in the position of the modes. Finally we provide a qualitative explanation for the dominance of one coherence over others. In Section \ref{sec:synch vs coherent transport} we demonstrate a correlation between coherent energy transfer and the time taken to reach positive synchronisation.

\subsection{Closed System Dynamics and the Conditions for Synchronisation} \label{sec: closed dynamics}
\begin{table*}[ht]
\centering
\begin{ruledtabular}
\begin{tabular}{ l | r r r r r r r }
& $|\psi_0\rangle\langle\psi_2|$ & $|\psi_0\rangle\langle\psi_3|$ & $|\psi_1\rangle\langle\psi_4|$ & $|\psi_1\rangle\langle\psi_5|$ & $|\psi_3\rangle\langle\psi_7|$ & $|\psi_3\rangle\langle\psi_8|$ & $|\psi_1\rangle\langle\psi_3|$
\\
\hline
$\Omega_{kj}$ (cm$^{-1}$) & 1111.0 & 1125.0 & 1102.6 & 1111.0 & 1111.0 & 1119.2 & 81.0
\\
$\langle\psi_k |X_1|\psi_j\rangle$ & 0.707 & -0.637 & 0.767 & 0.707 & 0.707 & -0.935 & -0.174
\\
$\langle\psi_k |X_2|\psi_j\rangle$ & 0.707 & 0.637 & -0.767 & 0.707 & 0.707 & 0.935 & 0.174
\\
$\langle\psi_k|\sigma_x|\psi_j\rangle$ & 0.000 & 0.385 & 0.340 & 0.000 & 0.000 & 0.384 & 0.196
\\
$\langle\psi_k|\Big(|0_1\rangle\langle 0_1|\otimes|0_2\rangle\langle 0_2|\Big)|\psi_j\rangle$ & 0.161 & -0.144 & -0.131 & 0.133 & 0.032 & 0.026 & -0.351
\end{tabular}
\end{ruledtabular}
\caption{Seven largest amplitude exciton-vibration coherences. Top row is the associated oscillation frequency from Equation \ref{eqn: components of X}. The remaining rows are the matrix elements corresponding to coherence $|\psi_j\rangle\langle \psi_k|$ of different operators which (in table order) represent: the coupling to position of mode 1, coupling to position of mode 2, coupling to inter-exciton coherence, coupling to ground state of both modes.}
\label{table of coherences}
\end{table*}

In order to understand the dynamical emergence of synchronisation between the positions of intramolecular modes we begin by exploring the time dependence of the expected values of the position operators, $\langle X_i(t)\rangle$.

In the basis of system eigenstates, the density matrix of the exciton-vibration systems reads $\rho(t) =\sum_{j,k} \rho_{jk}(t) |\psi_j\rangle\langle\psi_k|$ with $\rho_{jk}(t)=\langle \psi_j |\rho(t)|\psi_k\rangle$. The expectation value $\langle X_i(t)\rangle$ becomes
\begin{equation} \label{eqn: components of X}
    \begin{split}
    \langle X_i(t)\rangle & = \textrm{Tr}\big\{ X_i\rho(t)\big\} \\
     & = \sum_{l} \langle\psi_l|\bigg(X_i \sum_{j,k} \rho_{jk}(t) |\psi_j\rangle\langle \psi_k|\bigg)|\psi_l\rangle \\
     & = \sum_{j,k} \rho_{jk}(t) \langle \psi_k|X_i|\psi_j\rangle \\ & = \sum_{j,k} \rho_{jk}(t) X_{i,kj}
    \end{split}
\end{equation}
where $X_{i,kj} = \langle\psi_k|X_i|\psi_j\rangle$. Equation \ref{eqn: components of X} indicates that the time-evolution of local positions are given by the matrix elements $\rho_{jk}(t)$, yet the ability to exhibit synchronised behaviour depends critically on the form of $X_{i,kj}$ as it is the only source of difference between $\langle X_1(t)\rangle$ and $\langle X_2(t)\rangle$.

To explore the consequences of Equation \ref{eqn: components of X} in more detail we analyse whether synchronisation can occur in the closed quantum system whose dynamics is solely given by the system Hamiltonian $H$. Let us denote $\rho^H(t)$ as the density matrix of the closed system, which evolves according to:
\begin{equation} \label{eq: pjk closed}
    \begin{split}
    \rho^H(t) & = \sum_{j,k} \rho_{jk}(0)e^{i\Omega_{kj}t} |\psi_j\rangle\langle\psi_k|.
    \end{split}
\end{equation} 
Here $\rho_{jk}(0)=\langle \psi_j |\rho(0)|\psi_k\rangle$ are the populations ($j=k$) and coherences ($j\neq k$) of the initial state while $\Omega_{kj}=\epsilon_k-\epsilon_j$ where $\epsilon_j$ are the eigenenergies of $H$. One can therefore see that in a closed quantum dynamics only coherences will contribute oscillatory components, with specific frequencies $\Omega_{kj}$, to the dynamics of $\langle X_i(t)\rangle$.

To exhibit synchronisation we require oscillations in $\langle X_{1}(t)\rangle$ to align in frequency and phase with $\langle X_{2}(t)\rangle$ for an extended period of time. Using Equations \ref{eqn: components of X} and \ref{eq: pjk closed} this requires the equality:
\begin{equation} \label{eqn: synch equality}
    \sum_{j,k} \rho_{jk}(0) e^{i\Omega_{jk}t} X_{1,kj} = \sum_{j,k} \rho_{jk}(0) e^{i\Omega_{jk}t} X_{2,kj}
\end{equation}
for the oscillating components $j\neq k$. Equation \ref{eqn: synch equality} is trivially true if the initial state is an eigenstate of $H$. However this cannot be considered synchronised as there will be no time evolution at all. Hence Equation \ref{eqn: synch equality} can only be satisfied (and be non-zero) if the initial state $\rho(0)$ is such that element $\rho_{jk}(0)$ is zero when elements $X_{1,kj} \neq X_{2,kj}$. The only $\rho(0)$ that can fulfil these criteria contains either a single (or a specific combination of) eigenstate coherence(s) $|\psi_j\rangle \langle\psi_k|$.

\begin{figure*}
\subfloat[\label{cXX_vs_X closed}
]{\includegraphics[width=\textwidth]{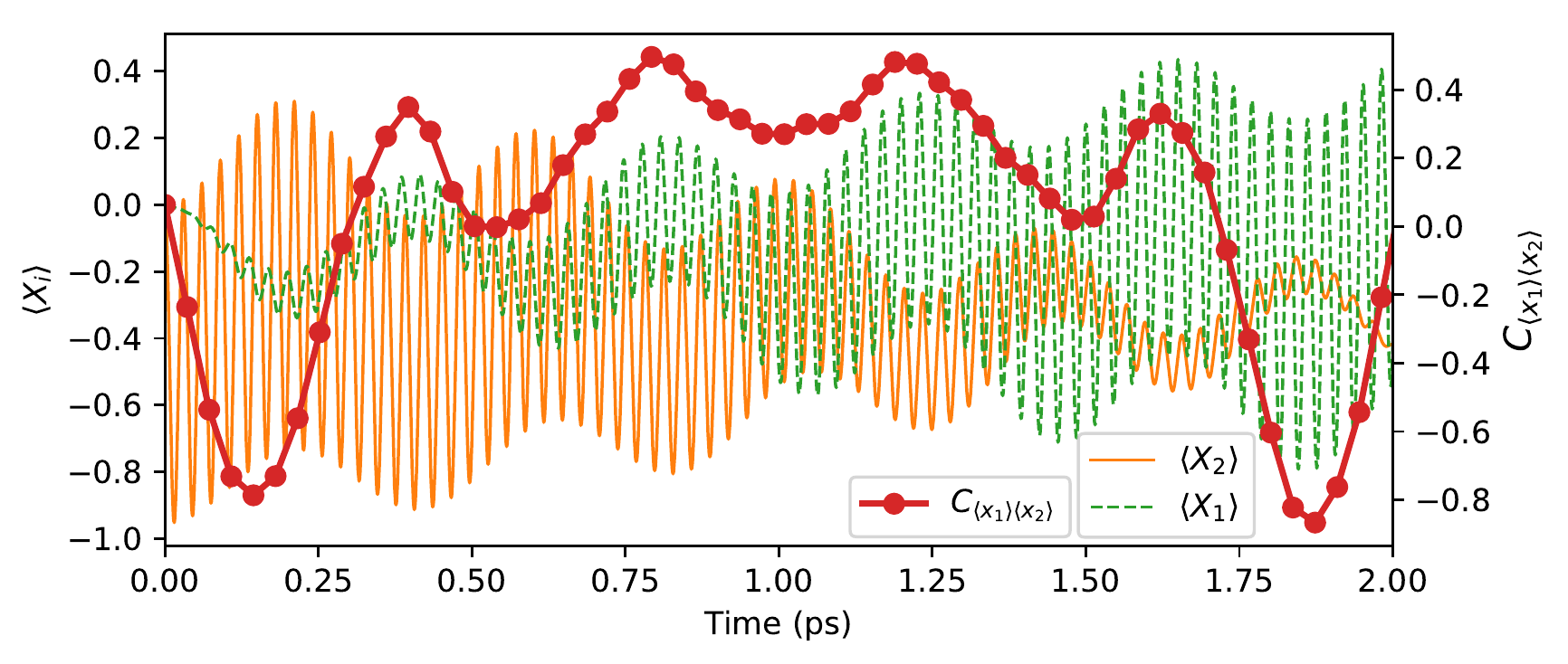}}\hfill
\subfloat[\label{FT closed}
]{\includegraphics[height=4.5cm]{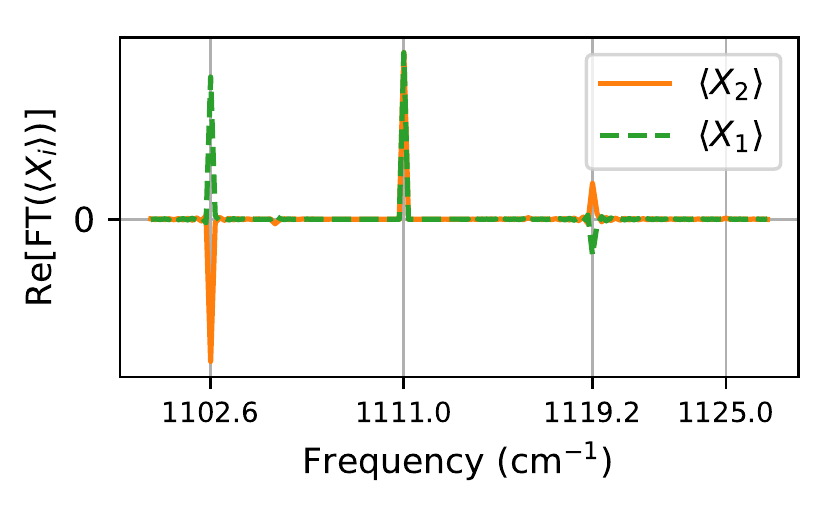}}\hfill
\subfloat[\label{Ex-vib coherence closed}
]{\includegraphics[height=6.7cm]{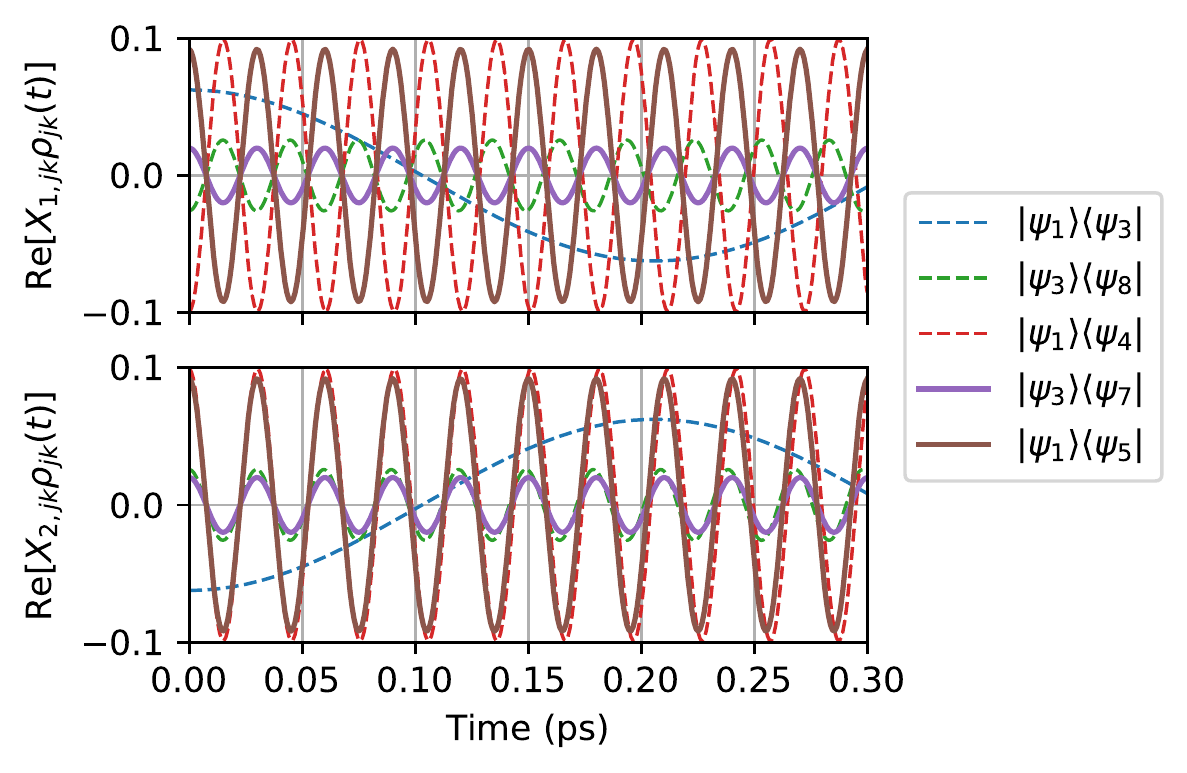}}\hfill
\caption{Coherent evolution of excited vibronic dimer as described in Section \ref{sec:methods}. Initial state is Equation \ref{eqn: initial state} and relevant parameters are listed in Table \ref{numerical parameters} (as this is closed system evolution the dissipation and dephasing rates are not applicable). (a) Evolution of expectation value of intramolecular mode positions $\langle X_i(t)\rangle$ and their synchronisation $C_{\langle X_1\rangle,\langle X_2\rangle}(t)$. (b) Real part of Fourier Transform of full time $\langle X_i(t)\rangle$ (c) Evolution of Real part of exciton-vibration coherences scaled by their coupling to $\langle X_1(t)\rangle$ [TOP] and $\langle X_2(t)\rangle$ [BOTTOM].}
\end{figure*}

For the system under consideration, we identify the seven largest position matrix elements  $X_{i,kj}$ contributing to the position dynamics and present their associated exciton-vibration coherences $|\psi_j\rangle\langle\psi_k|$ and frequencies, $\Omega_{kj}$, in Table \ref{table of coherences}. Examining these values shows that in the parameter regime considered (specifically $\omega_1=\omega_2$, see Section \ref{discussion} for more details) the position matrix elements fall into two distinct groups: those for which $X_{1,kj} = X_{2,kj}$ and those for which $X_{1,kj} = -X_{2,kj}$. For the latter set, the amplitude scaling of -1 results in a phase factor of $\pi$ between $\langle X_1(t)\rangle$ and $\langle X_2(t)\rangle$. For example a $\rho(0)$ consisting of $|\psi_1\rangle \langle\psi_5|$ and $|\psi_3\rangle \langle\psi_7|$ and no other coherences result in oscillations of frequency 1111.0 cm$^{-1}$ with no phase separation between $\langle X_1(t)\rangle$ and $\langle X_2(t)\rangle$. Similarly a combination of only $|\psi_1\rangle \langle\psi_4|$ and $|\psi_3\rangle \langle\psi_8|$ would result in oscillations at a frequency between 1102.6 cm$^{-1}$ and 1109.2 cm$^{-1}$ except with a constant phase separation of $\pi$. Both of these scenarios can be identified as synchronised as there is a constant phase difference over time: in the former the displacements are positively synchronised and in the latter they are negatively synchronised. A combination of the two groups of coherences however would create interferences yielding a cyclic phase change between $\langle X_i(t)\rangle$.

Whilst it is possible to find an initial state in which the dynamics of $\langle X_i(t)\rangle$ evolve in a synchronised way, this cannot be classed as spontaneous synchronisation as it has not emerged from an initially non-synchronised state. In the closed system evolution, the ratio of these coherences is determined only by the initial state $\rho(0)$ thereby fixing the frequency composition and precluding that there can be no dynamical emergence of synchronisation. The fact that synchronisation cannot emerge in the closed quantum system concurs with previous studies of coupled TLSs which show that synchronisation cannot occur in the presence of dephasing alone and some energy loss is required \cite{Giorgi2013}. This analysis allows us to postulate a mechanism for synchronisation in the open system which is as follows: In the presence of dissipation we would expect one (or a set) of coherences to emerge with a significantly larger amplitude than the others allowing it to dominate the dynamics of $\langle X_i(t)\rangle$ and produce a constant phase difference in $\langle X_i(t)\rangle$.

Figure \ref{cXX_vs_X closed} displays the numerical results of evolution of $\langle X_i(t)\rangle$ and the correlation measure $C_{\langle X_1\rangle\langle X_2\rangle}(t)$ in the closed system with initial state $\rho(0)$ (Equation \ref{eqn: initial state}). As expected, we see a large range of frequency oscillations in $\langle X_i(t)\rangle$. The value of $C_{\langle X_1\rangle\langle X_2\rangle}(t)$ changes in a cyclic pattern, indicating a continuous change in phase between the between the oscillations and a clear difference in frequency compositions. The frequency components in each $\langle X_i(t)\rangle$ can be resolved by taking the real part of the Fourier Transform (FT) which we present in Figure \ref{FT closed}. We note here how the frequencies present are exactly those in Table \ref{table of coherences} and that the frequencies that correspond to negative synchronisation can be clearly seen as those which have opposite sign in the FT.

Figure \ref{Ex-vib coherence closed} displays the short-time dynamics of the real parts of the five exciton-vibration coherences that dominate the evolution of $\langle X_i(t)\rangle$ (see Equation \ref{eqn: components of X}), weighted by their associated position matrix elements. Interference between these coherences generate the overall $\langle X_i(t)\rangle$ signals. Coherences $|\psi_1\rangle \langle\psi_5|$ and $|\psi_3\rangle \langle\psi_7|$ (bold lines) have identical frequencies and remain in phase throughout, whereas coherences $|\psi_1\rangle \langle\psi_4|$ and $|\psi_3\rangle \langle\psi_7|$ (dotted lines) begin to accumulate a phase difference due to their differing frequencies. The phase shift over time manifests as an oscillation in $\langle X_i(t)\rangle$ at a frequency equal to the differences between the pairs of $\Omega_{kj}$ involved. These are 8 cm$^{-1}$ (time period of 4.2 ps) and 17 cm$^{-1}$ (time period of 1.9 ps) which explains the approximate 2 ps periodicity seen in Figure \ref{cXX_vs_X closed}. The oscillation of period 0.4 ps is due to interference with low frequency coherence $|\psi_1\rangle \langle\psi_3|$ (dotted line slowly changing). It is clear that no single coherence dominates the dynamics and that synchronisation does not emerge from the chosen initial state in the closed system evolution.

\subsection{Open System Dynamics and the Emergence of Synchronisation} \label{Sec:Spontaneous Synch}

\begin{figure*}[ht]
\includegraphics[width=2\columnwidth]{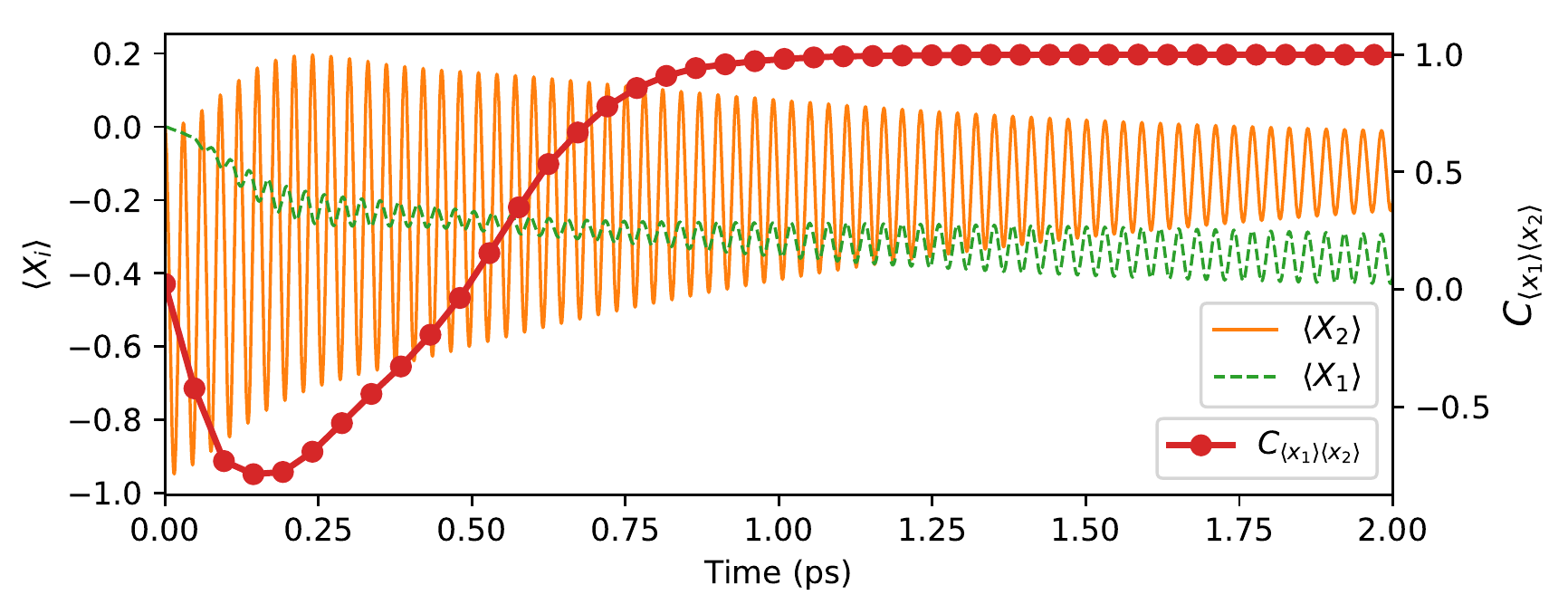}
\caption{Evolution of the expectation value of intramolecular mode positions $\langle X_i(t)\rangle$ and their synchronisation $C_{\langle X_1\rangle,\langle X_2\rangle}(t)$ in the open system evolution of an excited vibronic dimer as described in Section \ref{sec:methods}. Initial state is Equation \ref{eqn: initial state} and relevant parameters are listed in Table \ref{numerical parameters}}
\label{PE545:cXX_vs_X}
\end{figure*}

To understand how a synchronised state can emerge from a non-synchronised state we must understand the form of $\rho_{jk}(t)$ in the open quantum system. Solving Equation \ref{eq: master} for an element $jk$ results in a set of coupled differential equations: 
\begin{equation}
    \dot{\rho}_{jk}(t) = i\Omega_{jk}\rho_{jk}(t) + \sum_{\alpha,\alpha^\prime} R_{j,k,\alpha,\alpha^\prime} \rho_{\alpha,\alpha^\prime}(t)
    \label{eqn: redfield}
\end{equation}
where $R_{j,k,\alpha,\alpha^\prime}$ is commonly known as the Redfield tensor which captures the dependence of each $\rho_{j,k}$ on all other matrix elements as induced by the local dissipators. In this case we cannot say that each coherence $|\psi_j\rangle \langle\psi_k|(t)$ consists of only one single oscillation frequency $\Omega_{jk}$. However our numerical results show that the oscillatory dynamics of each of the seven coherences in Table \ref{table of coherences} are in fact dominated by the coherent component and that the latter terms in Equation \ref{eqn: redfield} contribute mainly a decaying dynamics. It is this decay that allows a change in ratio of coherences and the potential for synchronisation to emerge.

Figure \ref{PE545:cXX_vs_X} reports the synchronisation of $\langle X_1(t)\rangle$ and $\langle X_2(t)\rangle$ in the first two picoseconds of evolution. Inspection of the fast oscillations in the positions reveal an almost $\pi$ phase difference between $\langle X_i(t)\rangle$ at 0.15 ps and exactly in phase oscillations after 1 ps. This observation is captured quantitatively with $C_{\langle X_1\rangle,\langle X_2\rangle}(t)$ dipping to -0.75 towards negative synchronisation at 0.15 ps and then up to 1 for positive synchronisation at 1 ps. These numerical results show that synchronisation indeed occurs between the displacements of intramolecular modes of exciton-vibration dimers during the energy transfer process. To understand the underlying mechanism we perform an analysis similar to the previous section.

\begin{figure}[h]
\includegraphics[width=0.8\columnwidth]{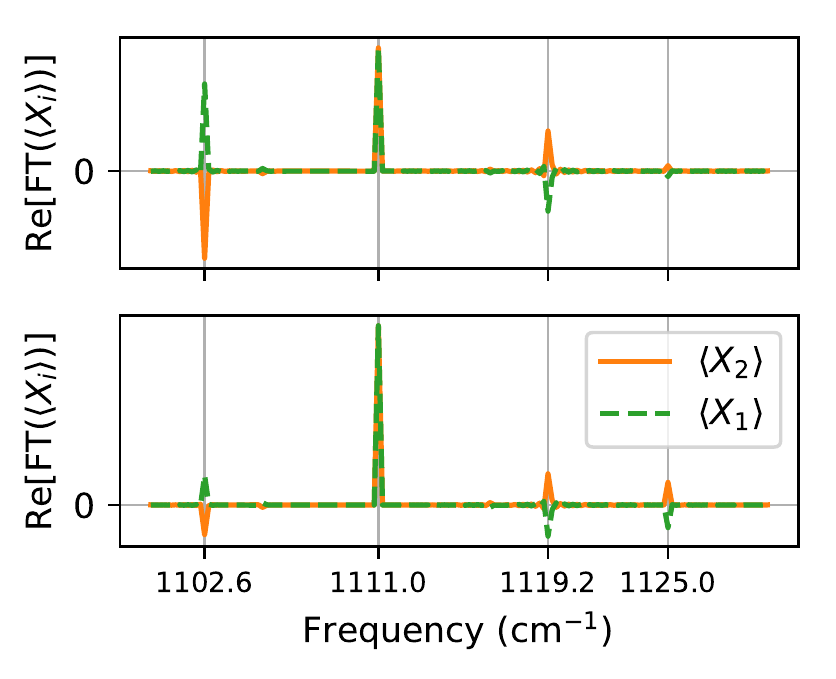}
\caption{Real part of Fourier Transform of $\langle X_i\rangle$ at 0.15 ps [TOP] and 1.50 ps [BOTTOM] during open system evolution.}
\label{PE545:FT}
\end{figure}

Figure \ref{PE545:FT} displays the FT of $\langle X_i(t)\rangle$ at 0.15 ps and 1.50 ps. Synchronisation can be seen again in this figure as the change in frequency distribution between the two time points. At 0.15 ps the FT resembles that of the coherent case in Figure \ref{FT closed} which indicates at this time coherent dynamics are dominating. The presence of negatively synchronised frequency 1102.6 cm$^{-1}$ at a magnitude comparable to positively synchronised frequency 1111.0 cm$^{-1}$ results in an interference and a non-stationary phase. This is reflected in $C_{\langle X_1\rangle,\langle X_2\rangle}(t)$ in Figure \ref{PE545:cXX_vs_X} at early times (0 - 1 ps) where the measure is continuously changing. At 1.5 ps we see the dominant frequency become 1111.0 cm$^{-1}$ which has equal amplitude in both $\langle X_i(t)\rangle$ signals and corresponds to the value of 1 in $C_{\langle X_1\rangle,\langle X_2\rangle}(t)$.

\begin{figure*}
\subfloat[\label{early}
]{\includegraphics[height=6.7cm,keepaspectratio]{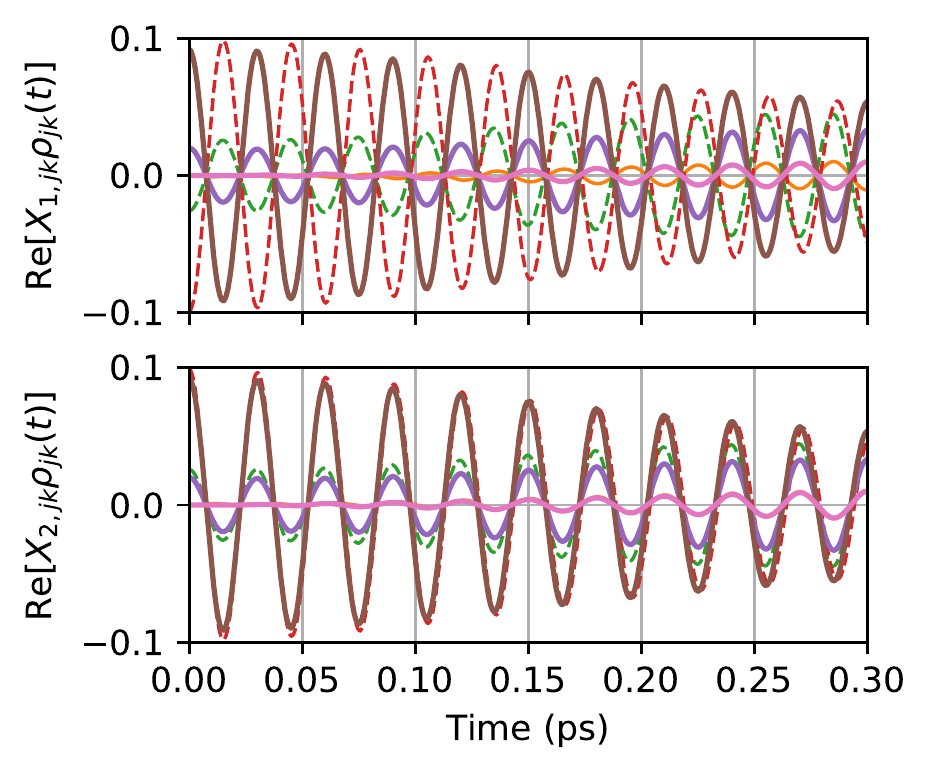}}
\subfloat[\label{late}
]{\includegraphics[height=6.7cm,keepaspectratio]{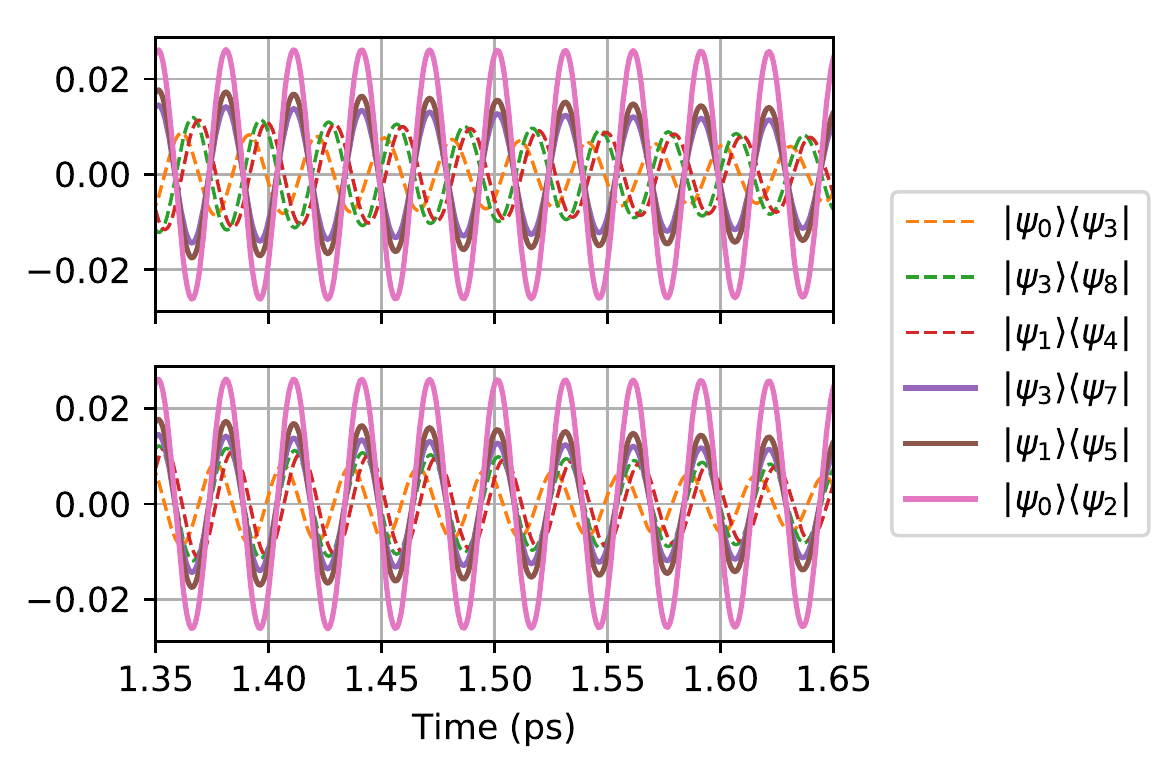}}
\caption{Evolution of Real part of exciton-vibration coherences in the open system scaled by their coupling to $\langle X_1(t)\rangle$ [TOP] and $\langle X_2(t)\rangle$ [BOTTOM] at times (a) 0 - 0.30 ps (b) 1.35 - 1.65 ps.}
\label{Fig: real coherences open}
\end{figure*}

Although it is useful for understanding the dynamics of synchronisation, this FT picture does not allow us to understand why the frequency composition of $\langle X_i(t)\rangle$ changes over time. To do so we must consider the underlying exciton-vibration coherence dynamics which we present in Figure \ref{Fig: real coherences open}. Initially we observe that the dominant frequency in the oscillation of each coherence is indeed the coherent part $\Omega_{jk}$ as can be evidenced by comparison to Figure \ref{Ex-vib coherence closed}. Figure \ref{early} presents the first 0.30 ps of evolution of coherences where the two signals are measured as being towards negatively synchronised. Figure \ref{late} presents the same coherences at 1.35 ps where they are measured as positively synchronised. We can immediately see how a change in the amplitude of the coherences has occurred in the latter case and the two signals appear much more similar. The dominant contributions come from the three positively synchronised coherences $|\psi_0\rangle \langle\psi_2|$, $|\psi_1\rangle \langle\psi_5|$ and $|\psi_3\rangle \langle\psi_7|$ (solid lines). These coherences constructively interfere to dominate the oscillations seen in $\langle X_i(t)\rangle$ and therefore a value of 1 in $C_{\langle X_1\rangle,\langle X_2\rangle}(t)$. It is this change in the ratio of coherences over time that determines the emergence of synchronisation.

\begin{figure}
\centering
\includegraphics[width=0.95\columnwidth]{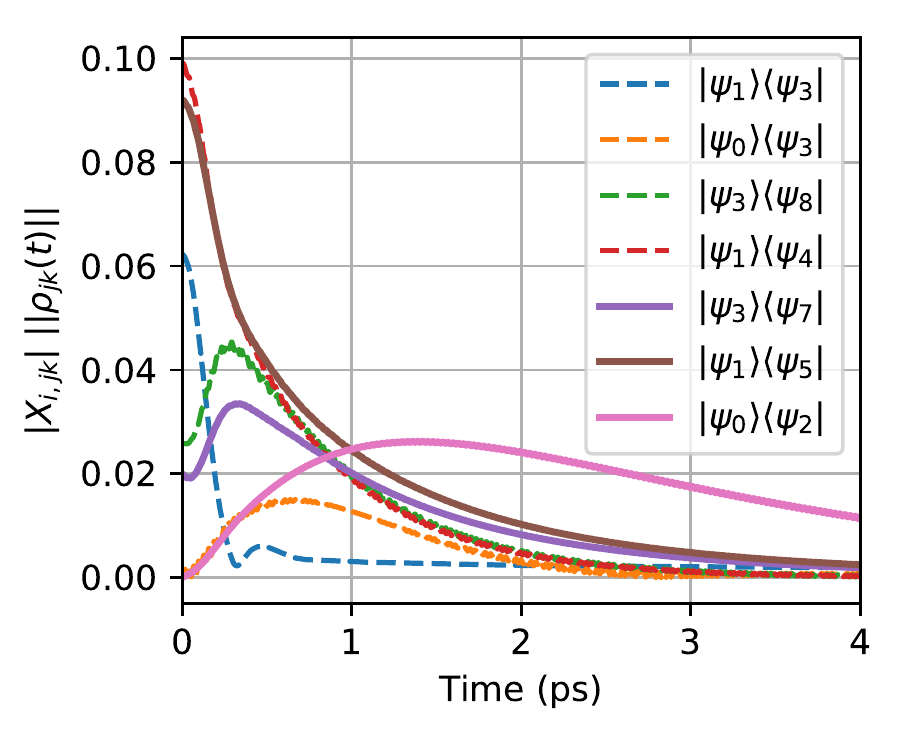}
\caption{Complex magnitude of exciton-vibration coherences scaled by the absolute value of their corresponding position matrix element in the open quantum system evolution.}
\label{PE545:Vibronic_coherences}
\end{figure}

In previous studies \cite{Giorgi2013} the mechanism for synchronisation has been related to a difference in the decay rates of eigenmodes of the Liouvillian superoperator, $\mathcal{L}$. The reasoning is that synchronisation occurs when one eigenmode of $\mathcal{L}$, that is equally coupled to the operators of interest such that their evolutions are synchronised, significantly outlives the other eigenmodes, transiently dominates the dynamics, and holds the operators in a synchronised state. Recently this explanation has been consolidated analytically with an exact treatment of a single dissipating qubit coupled to a probe qubit \cite{Giorgi2016}. In the original case these normal modes are found by diagonalising $\mathcal{L}$ and finding the conjugate pair of eigenvectors that have eigenvalues with real parts closest to zero and which couple significantly to the desired operator. Applying this process here results in an eigenmode of the Liouvillian that consists almost entirely of exciton-vibration coherence $|\psi_0\rangle \langle \psi_2|$. This analysis corroborates with our results which we present by plotting the absolute value of each coherence in Figure \ref{PE545:Vibronic_coherences}. We find coherence $|\psi_0\rangle \langle \psi_2|$ is indeed longest lived.

The Liouvillian eigenmode analysis provides a straight-forward prediction of the emergence of synchronisation at long times, but it does not facilitate an understanding of the early transient synchronisation dynamics. Neither does it provide understanding of why certain eigenmodes survive longer than others. The tracking of coherences we present in this paper however is capable of giving us insight into these early transients. In addition it allows us to give a qualitative explanation of why certain coherences survive longer in the presence of dissipation and dephasing.

We expect pure dephasing to result in exponential decays in excitonic coherences and thermal dissipation to cause exponential decays in the population of modes \cite{Breuer2002}. In the final two rows of Table \ref{table of coherences} we consider the matrix elements of an operator relating to excitonic coherences only $\sigma_x = |E_1\rangle\langle E_2|+|E_2\rangle\langle E_1|$, and the combined vibrational ground-state projector operator $|0_1\rangle\langle 0_1|\otimes |0_2\rangle\langle 0_2|$. Exciton-vibration coherences that result in a large matrix element for $\sigma_x$ will be more affected by the fast electronic dephasing. Similarly coherences that result in largest values for the projector of the ground vibrational states would last the longest due to thermal dissipation operating on a longer timescale and preferentially populating such ground states.

Out of the set of coherences considered in Table \ref{table of coherences} the one that has the lowest exciton coherence component and the largest ground-state vibrational component is $|\psi_0\rangle \langle \psi_2|$ which we find indeed to be the longest lived. This explanation can be consolidated with a numerical test in which we set the thermal dissipation rate faster than the dephasing rate (results not shown). In this case we would expect the eigenstate coherence containing a larger excitonic coherence component would survive longest. Indeed we find that the coherence $|\psi_0\rangle \langle \psi_1|$, which has a much larger excitonic coherence element $\langle \psi_0|\sigma_x|\psi_1\rangle = -0.8572$ would be the longest lived coherence and would lead to long lived negative synchronisation.

In summary we have shown how the interplay between exciton-vibration dynamics and the different noise sources considered lead to a rich synchronisation dynamics for the local modes and shown how said dynamics maps directly to the evolution of exciton-vibration coherences.

\subsection{The Role of Coherent Energy Transfer in Synchronisation of Intramolecular Modes} \label{sec:synch vs coherent transport}
\begin{figure*}
\subfloat[\label{synch_delocal}
]{\includegraphics[width=0.5\textwidth,height=6.3cm,keepaspectratio]{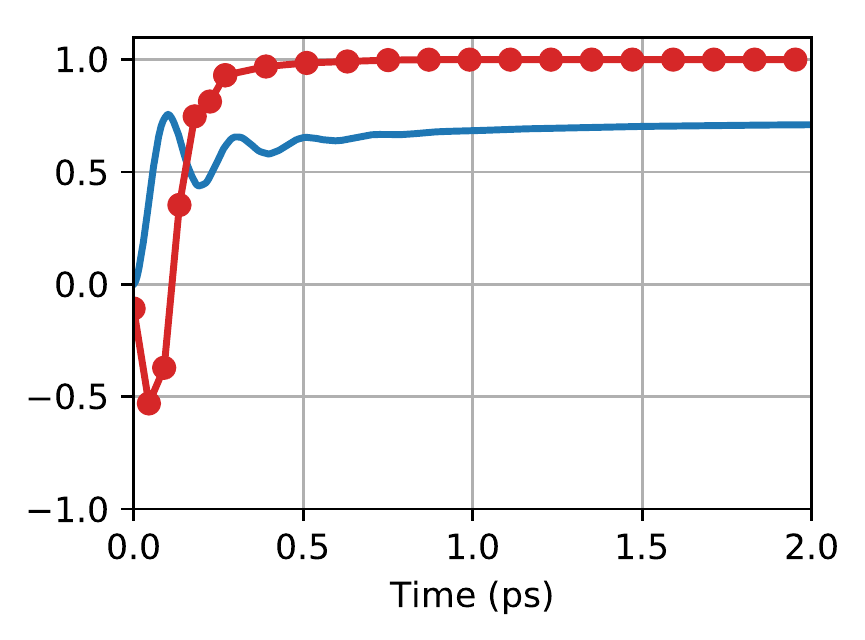}}\hfill
\subfloat[\label{coherences_delocal}
]{\includegraphics[width=0.5\textwidth,height=6.3cm,keepaspectratio]{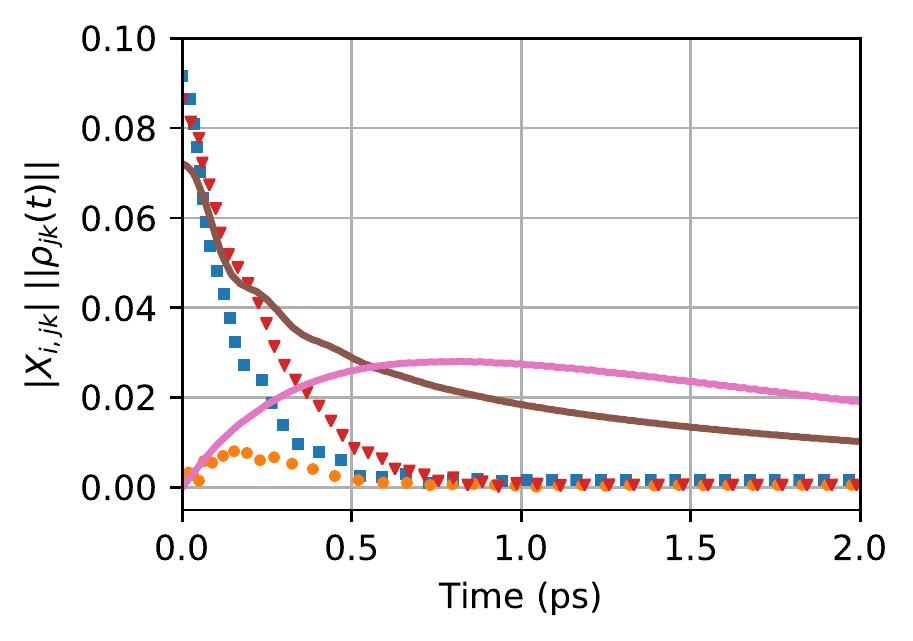}}\hfill
\subfloat[\label{synch_PE545}
]{\includegraphics[width=0.5\textwidth,height=6.3cm,keepaspectratio]{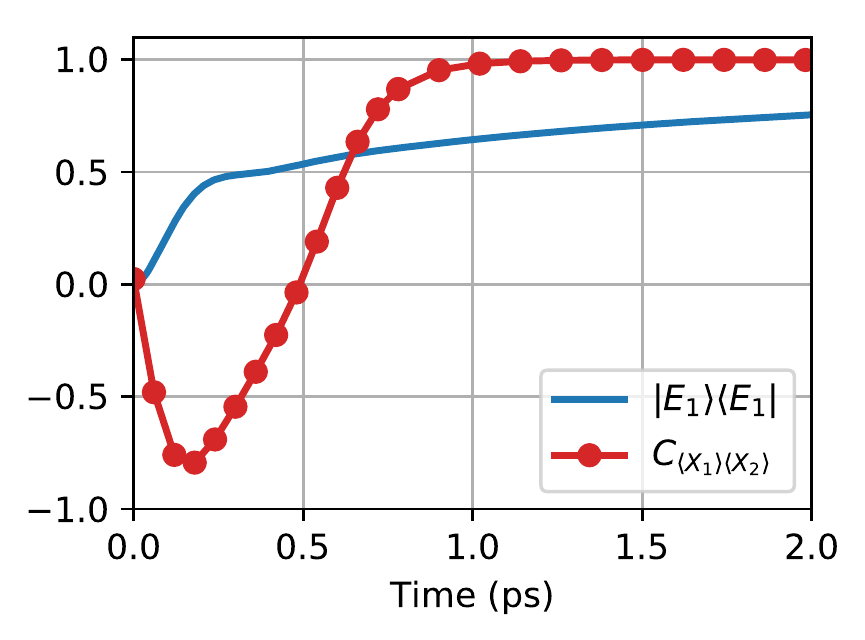}}\hfill
\subfloat[\label{coherences_PE545}
]{\includegraphics[width=0.5\textwidth,height=6.3cm,keepaspectratio]{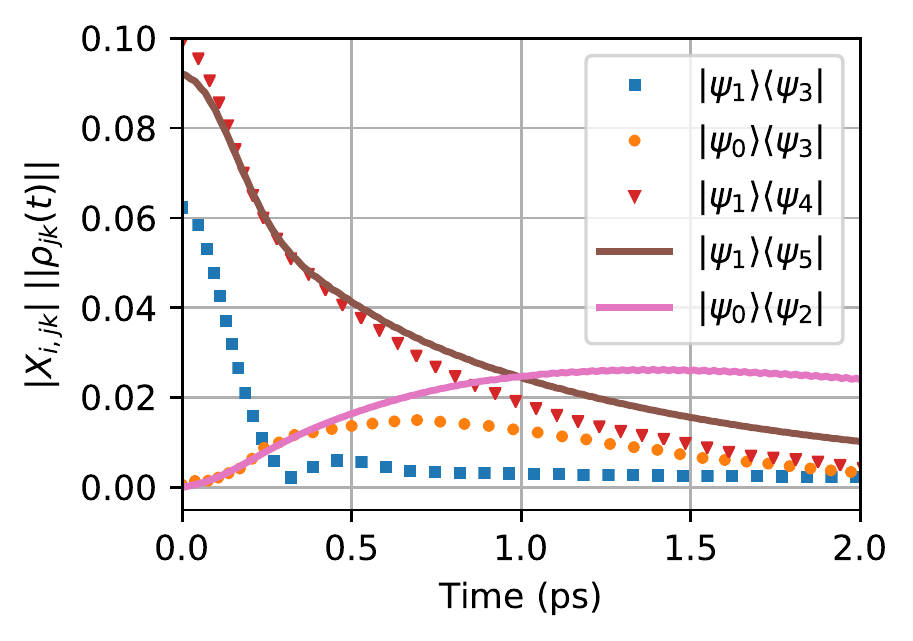}}\hfill
\subfloat[\label{synch_detuned}
]{\includegraphics[width=0.5\textwidth,height=6.3cm,keepaspectratio]{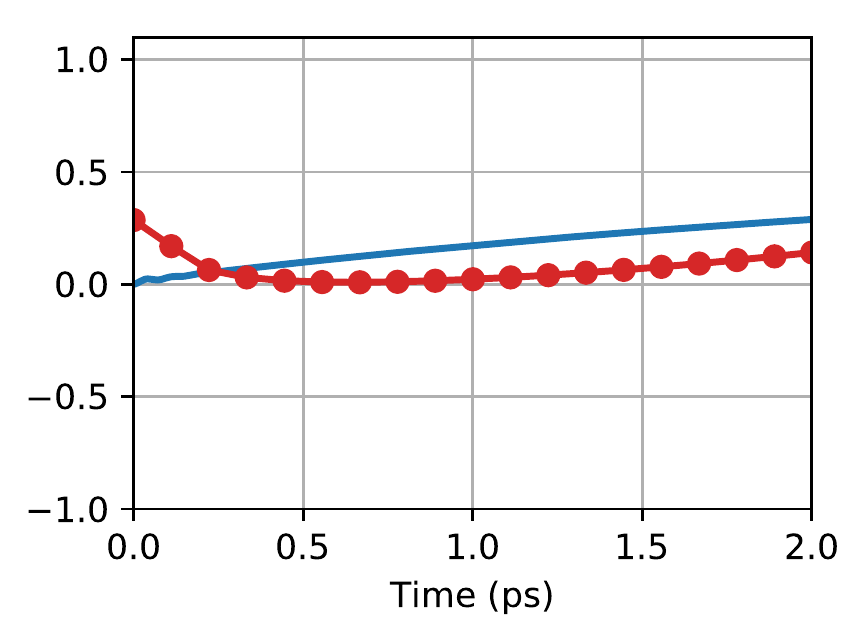}}\hfill
\subfloat[\label{coherences_detuned}
]{\includegraphics[width=0.5\textwidth,height=6.3cm,keepaspectratio]{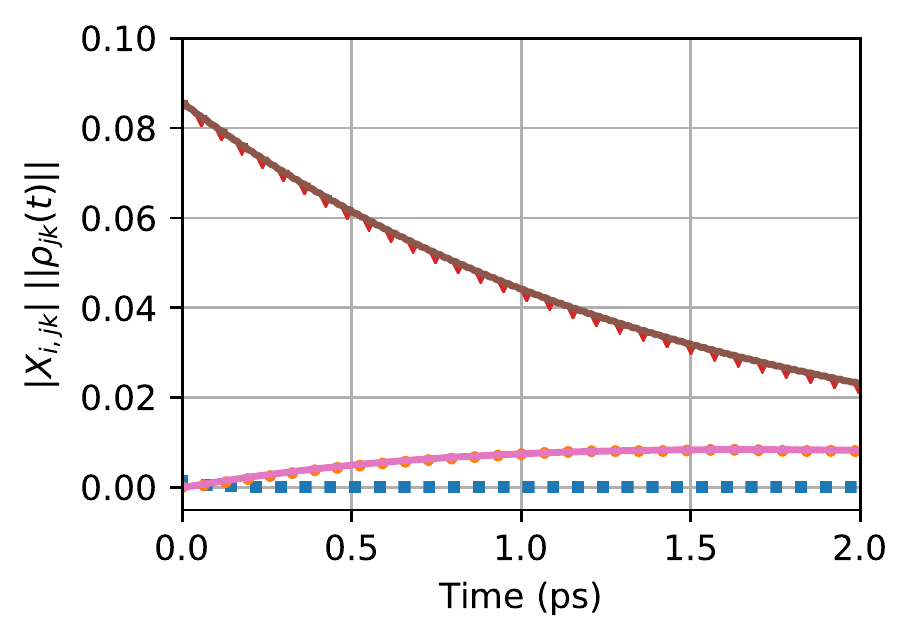}}\hfill
\caption{Comparison of dynamics of ET with synchronisation of $\langle X_i(t)\rangle$ and the magnitude of exciton-vibration coherences for three different parameter regimes: (c)(d) for central PEB dimer in PE545 (parameters in Table \ref{numerical parameters}) corresponding to A=0.76; (a)(b) modified parameters $\eta = 0.5$ corresponding to A=0.95; (e)(f) modified parameters $\omega = 1500\mbox{ cm}^{-1}$ corresponding to A=0.04.}
\label{Gridfig}
\end{figure*}
Thus far we have described the coherent and dissipative contributions to the dynamics of exciton-vibration coherences and therefore to the dynamics of synchronisation. Presently we turn our attention to the electronic energy transfer dynamics of the light-harvesting unit considered, its coherent character and its relation to synchronisation of local displacements.

One of the most important features of the prototype system considered is that efficient ET is aided by a resonance in energy between the exciton energy splitting and an energy quanta of the local intramolecular modes \cite{OReilly2014, Kolli2012, Dean2016}. This resonance results in a range of eigenstates $|\psi_j\rangle$ that are close in energy yet in the quasi-local basis of $|E_d, n_1, n_2\rangle$ have significantly different weights. When coherently evolving from the initial state considered (with the excitonic system being initially in the highest-energy exciton state) joint exciton-vibrational transfer pathways are open and population can be coherently transferred to eigenstates involving the lowest lying exciton.

We can intuitively understand how ET is essential for the synchronisation of local displacements in the prototype dimers studied by recalling both the quasi-localised nature of the excitons for the parameter regime considered and the local nature of the electronic-vibration interactions. Trivially, ET between excitons must occur for the intramolecular modes to become effectively coupled, exchange energy and synchronise. However the precise relations between the degree of coherent electronic ET and synchronisation is less obvious. In our synchronisation analysis so far we observe that the ET period (0 - 0.5 ps) is concomitant with the negatively synchronised transient of Figure \ref{PE545:cXX_vs_X}. It appears that during this energy transfer period the displacements of the modes tend towards being negatively synchronised. This suggests a signature of coherent ET could be found in a measure of synchronisation of local mode displacements, which we discuss further in Section \ref{discussion}.

To investigate the relationship between coherent ET and synchronisation quantitatively, we compare scenarios in which the resonance condition between exciton energy splitting and mode energy quanta is kept fixed but the degree of delocalisation of excitons is increased such that coherent ET transfer is enhanced. We also analyse the case in which the frequency of the modes are detuned from the exciton energy splitting to illustrate the fundamental role of the energy matching condition both for ET and for synchronisation.

The energy difference between exciton splitting and vibrational energies, i.e. $\Delta=\Delta E-\omega$; the coupling strength between local vibrational and electronic degrees of freedom $g$; and the exciton size or delocalisation $\theta$ which depends on the ratio $\eta =2|V|/\Delta E$ all influence the coherent character of ET. An approximate indicator of the degree of coherent ET is derived\cite{OReillythesis} from the transition probability between the two exciton-vibration states dominating electronic ET in our prototype dimer. The indicator is an estimate of the maximum amplitude $A$ for the population oscillations and is given by:
\begin{equation}
A = \frac{1}{1+(\frac{\Delta}{2g\sin(2\theta)})^2}.
\end{equation}

The parameter regime used throughout this paper (see Section \ref{sec:methods}) corresponds to $A=0.76$. For the off-resonant modes we choose a frequency of 1500 cm$^{-1}$ (a mode at this frequency is also present in PE545 \cite{Viani2014}), resulting in $A=0.04$. For increased exciton delocalisation we choose $\eta = 0.5$ but keep the energy resonance condition $\Delta$ fixed, resulting in $A=0.95$. In Figure \ref{Gridfig} we present the lowest exciton population dynamics (operator $|E_1 \rangle \langle E_1|$), which provides numerical evidence of the change in ET, alongside the dynamics of synchronisation and the selected exciton-vibration coherences (see Figure \ref{PE545:Vibronic_coherences} from the previous analysis) in each scenario. Together these plots allow us to compare the three different ET regimes and consider their effects on synchronisation.

Firstly we note the clear change in magnitude of coherent ET from high in Figure \ref{synch_delocal}, to low in Figure \ref{synch_detuned} and that the time taken for synchronisation to occur appears to follow the same pattern. Positive synchronisation is achieved in 0.5 ps when $A=0.95$, 1 ps when $A=0.76$ and is not reached in the 2 ps window presented when $A=0.04$. This correlation is also reflected in the exciton-vibration coherence evolutions in Figures \ref{coherences_delocal}, \ref{coherences_PE545} and \ref{coherences_detuned} (note we have removed two of the seven exciton-vibration coherences for clarity). When $A=0.95$ (Figure \ref{coherences_delocal}) negatively synchronised coherences (plotted with shape markers) decay faster than when $A=0.76$ (Figure \ref{coherences_PE545}). This results in the positively synchronised coherences (solid lines) dominating from an earlier time and hence the earlier emergence of positive synchronisation. When $A=0.04$ (Figure \ref{coherences_detuned}) we observe that positively and negatively synchronised coherences have almost identical amplitude throughout the evolution and therefore neither can dominate. Small differences are amplified and interferences prevent synchronisation in the time scale considered. Positive synchronisation only emerges at around 10 ps (not shown) when positive synchronised coherences finally outlive the negative synchronised ones however by this time the oscillation amplitudes have nearly decayed to zero.

Additionally we identify that the eigenstate coherence $|\psi_1\rangle\langle\psi_3|$ can be used as an indicator of the magnitude of coherent ET. In the three regimes considered its amplitude appears correlated with the amplitude of coherent ET. This can be understood by considering the composition of the eigenstates involved and their energy splitting. Firstly, the eigenstates $|\psi_1\rangle$ and $|\psi_3\rangle$ have an energy splitting of only 81 cm$^{-1}\ll \omega$ which leads to resonance energy transfer between them. Secondly, in the basis of $|E_d, n_1, n_2 \rangle$, these eigenstates have the following approximate compositions:
\begin{equation}
\begin{split}
|\psi_1\rangle & \approx 0.3|E_1\rangle \left(|01\rangle - |10\rangle \right) +0.9|E_200\rangle - 0.2|E_201\rangle, \\
|\psi_3\rangle & \approx 0.6|E_1\rangle \left(|01\rangle - |10\rangle \right) -0.4|E_200\rangle +0.1|E_201\rangle.
\end{split}
\end{equation}
which shows that a transition between them would transfer exciton population. Together this suggests the coherence $|\psi_1\rangle\langle\psi_3|$ is part of the ET mechanism and explains why its amplitude scales in accordance with the changes in ET.

Our analyses clearly highlight a correlation between the amplitude of coherent excitation transport and the time it takes for mode displacements to synchronise. This is significant as it suggests that a degree of control of quantum synchronisation can be achieved by adjusting only coherent exciton-vibration interactions and specifically without changing the environment-induced dissipation or dephasing.

The relationship we observe can be understood as follows: parameter changes that increase ET effectively increase the coupling between the subsystems and allow a faster exchange of energy; the electronic subsystem `overshoots' its equilibrium position and ET oscillations occur; the damping of pure exciton coherences accelerates and in turn the decay of negatively synchronised exciton-vibration coherences accelerates; the system becomes dominated by positive synchronised coherences in a shorter time scale and exhibits the shorter synchronisation times we observe.

\section{Discussions and Conclusion} \label{discussion}
Before concluding we discuss some of the implications and limitations of our results.

The results of Sections \ref{Sec:Spontaneous Synch} and \ref{sec:synch vs coherent transport}, indicate negative synchronisation is concomitant with coherent energy transport. Moreover, our analysis of the exciton-vibration coherences that are negatively synchronised showed that they all have a large component of excitonic coherence. This raises the question: can we conclude that the negative synchronisation of local mode displacements is a signature of the survival of excitonic coherence? A transient shift towards negative synchronisation persists in all different initial preparations of exciton state (except for an initial state equal to the the steady state), and the shift becomes less pronounced as coherent exciton population transport inhibited. Additionally, as reported in Section \ref{Sec:Spontaneous Synch}, if dephasing rates are much slower than thermal relaxation i.e. excitonic coherence is longer lived, we find that the length of this negative synchronisation period extends to the steady state. Altogether this suggests that, for the system considered, a negatively synchronised transient in the positions of intramolecular modes can indeed be a signature of excitonic coherence and coherent energy transfer.

The parameter regime studied throughout this paper is for the special case of $\omega_1 = \omega_2$. This restriction allowed us to focus on the complex relationship between synchronisation, coherence and dissipation. However it also raises some interesting points.

Firstly, a classical view of two coupled oscillators with $\omega_1 = \omega_2$ might lead one to expect synchronisation to trivially always occur. We have shown that in the quantum setting, due to the exciton-vibration nature of the complex, the frequencies at which the local mode positions may oscillate are not equal and, in fact, change over time. This effect cannot be thought of classically. On the other hand, the relationship between ET and synchronisation can be thought of as analogous to how increasing the coupling strength between two classical oscillators allows them to synchronise faster. Our intramolecular modes are coupled to local electronic states that exchange energy through electronic coupling. Increasing ET involves increasing the electronic coupling which leads to a stronger effective coupling between the modes.

Secondly, as mentioned in Section \ref{Sec:Spontaneous Synch}, the exciton-vibration coherences we study have precisely equal or opposite values in $X_{i,kj}$. We can explain this result by examining our Hamiltonian for differences between $X_1$ and $X_2$. Whilst $\omega_1=\omega_2$, swapping $X_i$ would result in no changes to $H_{vib}$ and a sign change on excitonic coherences in $H_{exc-vib}$ i.e. $\langle E_2|\Theta_1|E_1\rangle = -\langle E_2|\Theta_2|E_1\rangle$. This -1 scaling is the source of the differences between $X_{1,kj}$ and $X_{2,kj}$ seen in this investigation. If $\omega_1\neq\omega_2$, $H_{vib}$ is no longer symmetric upon exchange of $X_i$. We expect this detuning would result in a wide range of $X_{i,kj}$ and therefore contributions of different phases to the dynamics.

To conclude, we have predicted the transient spontaneous synchronisation of the displacements of intramolecular modes on neighbouring molecules in a bio-inspired vibronic dimer. Until now, synchronisation had not been investigated in a hybrid quantum system where excitonic and vibrational coherence overlap in such a way. We have presented an understanding of the mechanism for synchronisation as the survival of specific exciton-vibration coherences, and detailed how coherences are selected for by dissipation. This analysis may provide a perspective from which we can understand synchronisation in other dissipating hybrid quantum systems such as larger multi-chromophore systems. We showed that both dissipative and coherent dynamics play an important role in the formation of synchronisation in these systems. Coherent ET is positively correlated with the time scale in which synchronisation is achieved. Dissipation is required for the decay of exciton-vibration coherences to allow one to dominate and can control the form of synchronisation that occurs. The parameter regime that most closely resembles a real photosynthetic dimer appears to be one where there is a balance between coherent and dissipative dynamics in which synchronisation can emerge before the steady state is reached. Our work highlights a novel possible role for exciton-vibrational coherence in biomolecular complexes, namely supporting the synchronisation of out-of-equilibrium vibrational motions.

\bibliography{Mendeley}
\end{document}